\documentclass[letterpaper, 10 pt, conference]{ieeeconf}  

\IEEEoverridecommandlockouts                              
\makeatletter
\let\NAT@parse\undefined
\makeatother

\usepackage[dvipsnames]{xcolor}

\newcommand*\linkcolours{ForestGreen}

\usepackage{times}
\usepackage{graphics}
\usepackage{graphicx}
\usepackage{amssymb}
\usepackage{gensymb}
\usepackage{amsmath}
\usepackage{breakurl}

\usepackage{url,hyperref}
\hypersetup{
colorlinks,
linkcolor=\linkcolours,
citecolor=\linkcolours,
filecolor=\linkcolours,
urlcolor=\linkcolours}

\usepackage{algorithmic}
\usepackage{algorithm2e}
\usepackage[labelfont={bf},font=small]{caption}
\usepackage[none]{hyphenat}

\usepackage{mathtools, cuted}

\usepackage[noadjust, nobreak]{cite}

\usepackage{tabularx}
\usepackage{amsmath}

\usepackage{float}

\usepackage{pifont}

\newcolumntype{Y}{>{\centering\arraybackslash}X}

\usepackage[normalem]{ulem} 

\usepackage[]{placeins}

\usepackage{placeins}
\usepackage{caption}
\usepackage{tikz}

\usepackage[framemethod=tikz]{mdframed}

\usepackage{afterpage}

\usepackage{stfloats}

\usepackage{atbegshi}
\newcommand{\handlethispage}{}
\newcommand{\discardpagesfromhere}{\let\handlethispage\AtBeginShipoutDiscard}
\newcommand{\keeppagesfromhere}{\let\handlethispage\relax}
\AtBeginShipout{\handlethispage}

\usepackage{comment}
\usepackage{lipsum}

\title{\LARGE \bf
Controller-based Energy-Aware Wireless Sensor Network Routing using Quantum Algorithms
}
\usepackage[textsize=footnotesize,backgroundcolor=red!25]{todonotes}

\author{Jie Chen, ~Prasanna Date, ~ Nicholas Chancellor, ~  Mohammed Atiquzzaman, ~Cormac Sreenan	
\thanks{Jie Chen is a research associate in the physics department of Durham University (e-mail:jie.chen@durham.ac.uk)}
\thanks{Prasanna Date is a research scientist of Oak Ridge National Laboratory (e-mail: datepa@ornl.gov)}
\thanks{Nicholas Chancellor is a research fellow in the physics department of Durham University (e-mail: nicholas.chancellor@durham.ac.uk)}
\thanks{Mohammed Atiquzzaman is a professor in the computer science department of Oklahoma University (e-mail:atiq@ou.edu)}
\thanks{Cormac Sreenan is a professor of computer science at University College Cork (e-mail:cjs@cs.ucc.ie)}
}

\begin{document}

\maketitle
\thispagestyle{empty}
\pagestyle{empty}

\begin{abstract}
Energy efficient routing in wireless sensor networks has attracted attention from researchers in both academia and industry, most recently motivated by the opportunity to use SDN (software defined network)-inspired 
 approaches. These problems are NP-hard, with algorithms needing computation time which scales faster than polynomial in the problem size. Consequently, heuristic algorithms are used in practice, which are unable to guarantee optimally. In this short paper, we show proof-of-principle for the use of a quantum annealing processor instead of a classical processor, to find optimal or near-optimal solutions very quickly. Our preliminary results for small networks show that this approach using quantum computing has great promise and may open the door for other significant improvements in the efficacy of network algorithms.
\end{abstract}

\section{Introduction}
Wireless sensor and ad-hoc networks have attracted much attention in research recently, including the Internet of Things (IoT), embedded systems and autonomous vehicles. The network structure is usually non-hierarchical and autonomously forms a network faciliated by ad-hoc mobile network protocols. In order to achieve resource utilization efficiency in terms of energy consumption, the concept of a network controller device is seen as appropriate\cite{access}, influenced by software defined networking (SDN)\cite{sdwmnl}. Using SDN at the network edge with devices that are resource-constrained, wireless and perhaps mobile, raises many challenges, but presents a great opportunity to use algorithms for routing and other tasks that can take advantage of centralised computation. In our case, the network controller uses a hybrid of classical and quantum algorithms to minimize the energy consumption and also meet the expected average data rate metrics for the network. The controller runs the computation of the optimal path set for a given set of packet streams at various inception nodes and assigns the path to each stream. 

\subsection{Background}
Quantum computers use quantum mechanical phenomena of superposition, entanglement and tunneling to perform computation.
They operate in the tensor-product Hilbert spaces and can perform computations in exponentially large dimensions to solve complex problems. 
Quantum algorithms are known to outperform classical algorithms on many challenging problems such as integer factoring \cite{shor1994algorithms}, search \cite{grover1996fast}, Fourier transform \cite{coppersmith2002approximate} and training machine learning  models \cite{date2020adiabatic,date2021qubo}. 
Several quantum and quantum-classical hybrid algorithms have also been proposed to address different types of optimization problems. 
For instance, the HHL algorithm for least square fitting \cite{harrow2009quantum}, quantum semidefinite programming algorithm for semidefinite programming \cite{brandao2017quantum}, quantum approximate optimization algorithm for combinatorial optimization \cite{farhi2014quantum}, adiabatic quantum computing for quadratic programming as well as NP-complete problems \cite{arthur2020balanced} and variational quantum eigensolver for nonlinear optimization \cite{peruzzo2014variational}.

In this work, we focus on the D-Wave adiabatic quantum computers, which were designed to approximately solve the quadratic unconstrained binary optimization (QUBO) problem, which is NP-hard.
As such, these machines have been used to address many NP-hard problems in the literature in purely quantum as well as quantum-classical hybrid approaches. 
D-Wave machines have been used to address the graph partitioning \cite{mniszewski2016graph,ushijima2017graph} and graph coloring problems \cite{wikeckowski2019disorder}.
Warren addresses the traveling salesman problem using the D-Wave quantum annealing computers \cite{warren2020solving}.
Date et al. propose a quantum-classical hybrid algorithm to train restricted Boltzmann machines and deep belief networks \cite{date2019classical}.
One of the biggest challenges pertaining to adiabatic quantum computing is the embedding a given QUBO problem onto the hardware of the quantum computer \cite{date2019efficiently}.
In this work, we first convert the energy efficient network routing problem as a QUBO problem and then leverage two D-Wave processors, both 2000Q and Advantage to solve it, we find that both can provide faster (in terms of raw processing time) solutions than state-of-the-art-classical solvers for some small network problems. While these problems are too small to directly show a quantum advantage, these are still hopeful signs in terms of the ability of these devices to find a ``good solution quickly''.

\subsection{Novelty and Contribution}
This work has made three breakthroughs in the field of quantum computing, being the first to:
\begin{itemize}
    \item Engage the computation power of a QPU to network design, in this instance focused on energy management.
    \item Compare the 2000Q and Advantage\_System1.1 processors in solving a network design problem. 
    \item Apply the Domain Wall Encoding scheme \cite{nick} for QPU in a practical engineering problem.
\end{itemize}
The work demonstrates the merits of applying a quantum annealing QPU in network design, but also provides a simulation platform that can be used by other researchers for future QPU-facilitated design and test problems.

We start by discussing the current state of the art and related work which we build upon. We then discuss the general formulation of the problem by conventional methods and how to translate it to the recently proposed domain-wall encoding, and propose an algorithm for mapping the problem. We then describe the details of our experimental methods and report our results. Finally we discuss the consequences of the results and the longer term outlook. 

\section{Related Work}
\label{sec:related-work}
Energy consumption in network routing is caused by neighbourhood discovery, communication and computation. Energy efficient routing with quality of service (QoS) guarantee in different applications or diverse wireless sensor networks (WSNs) can be viewed as an interesting area for future investigation \cite{access}. Unbalanced energy consumption among the nodes causes network partition and node failures, where transmission from some nodes to the sink becomes blocked \cite{javaideewsn}. Energy efficiency can be improved at various layers of the communication protocol stack of WSN. For hardware-related energy efficiency, topics have been focusing on lower power electronics, power-off mode and energy efficient modulation. For network-layer related energy Lee et.~al.~\cite{vwmn} proposed the routing schemes in consideration of both the link quality and the residual energy level. It discusses the mechanism for forwarding route requesting packets by calculation of the probability to forward or not by taking into account the link quality and residual energy level two metrics. The scheme is evaluated against the plain AODV (Ad hoc On-Demand Distance Vector) algorithm. In \cite{straightlinewsn}, Liu, Su, and Chou presented a simple and highly efficient strategy to form the energy aware path from source to sink node in wireless sensor network. When the event path and query path intersects, an anchor node is discovered or it can be found within the candidate region around the intersection point. It solves the spiral problem in rumour routing by keeping the event and query paths as straight as possible and proves to outperform rumor routing and achieves higher successful path discovery ratios and lower hop counts and saves more energy. In \cite{DelayLossyN} Zhao et.~al.~solve the optimal routing path issue of wireless sensor network by formulating the problem using the optimal path set which consists of possible combination of nodes that falls within the optimal transmission range of the source node and have minimum cost value corresponding to the cost function with energy and loss rate as the input parameters. The author firstly derives the metrics to evaluate the cost function and then proposes a brute force scheme to scan all the possible combination of nodes to determine the optimal path set. Experiments have shown that the scheme can provide robust connectivity and prolong the lifetime of the network compared with benchmarks across different scenarios.
Yao, Cao, and Vasilakos \cite{heternetwork} solve the energy efficiency problem within wireless sensor network while not violating QoS metrics by formulating the problem in the framework of the open vehicle routing problem in operation research. As the OVR (open vehicle routing) problem proves to be NP-hard, the authors subsequently proposed two different heuristic algorithms to approximate the design outcome. The outputs generated demonstrated to outperform baseline protocols in achieving longation of the network lifetime within the expected delivery latency bounds.Due to complexity all today's network routing solutions for WSNs rely on heuristic algorithms

In this work, the computation for the optimal set of paths is achieved by formulating the problem as a quadratic unconstrained binary optimization (QUBO) problem that can be easily mapped to the Ising model.QUBO has been used to solve network routing and similar optimisation problems in literature \cite{furqan,siya,irie} and it is just one of possible formulation selected to demonstrate the power of quantum.
The optimization can be solved by applying the quantum annealing technique used in physics to attain the ground state of the final state of the Hamiltonian system specified by the Ising model co-efficients. We are using the Quantum Processor Unit (QPU) by D-Wave Systems Inc in this work.

\section{Problem Formulation \label{sec:prob_form}}
\begin{table*}[ht]
\caption{The list of symbols used in this paper}
\centering
\begin{tabular}{p{0.15\linewidth}p{0.85\linewidth}}
\hline
Symbol & Definition \\
\hline
$\Delta t$       &   network controller monitoring interval \\
  $t_{org}$       &  the absolute starting time of the network controller   \\
  $s_i$       & the $i^{th}$ source node  \\
  $d$      & the sink(destination) node   \\
  $p_j$       & the $j^{th}$ packet stream   \\
  $r_j$ & the average data rate for the $j^{th}$ packet stream \\
  $(s_{i,j},d)$       & the pair of the $j^{th}$ packet stream starting from the $i^{th}$ source node to the sink    \\
  $route_{j,k}$       & the $k^{th}$ path for the $j^{th}$ packet stream    \\
  $edge_m$       & the $m^{th}$ edge \\
  $K$       & the maximum number of paths available for every packet stream    \\
  $l_j$       &  the length of the $j^{th}$ edge   \\
  $p_{i,j}$       & the $j^{th}$ path for packet stream starting from the $i^{th}$ source node\\
  $edg_i$ & the $i^{th}$ edge\\
  $\bar{X_i}$ & the indicator vector to describe which path is selected for packet stream starting from source node $i$\\
  ${xp}_{i,j}$ & the $j^{th}$ path for the packet stream starting from source node $i$. 1: the path is selected;0: the path is not selected\\
  $e_{i,j}$ & the energy consumption for packet stream starting from source node $i$ and travel through path $j$\\
  $e_{i,j,k}$ & the energy consumption at $k^{th}$ edge with source number $i$ whose path index is $j$\\
  $C_{max}$ & maximum link rate per edge\\
  $E_{Tx}(l,d)$ & the transmitter energy consumption of $l$ bits of data across distance $d$ \\
  $E_{Rx}(l,d)$ & the receiver energy consumption of $l$ bits of data across distance $d$\\
  $f(d_j)$ & the transmission energy consumption for the $j^{th}$ edge with distance $d$\\
  $L$ & the number of packet streams\\
  $\bold{I}_{i,j}$ & the indicator function to indicate whether the $j^{th}$ edge is selected for packet stream starting from node $i$ or not\\
\hline
\end{tabular}
\end{table*}

For this problem, we assume the network controller monitors the data traffic at a regular basis $\Delta t$, starting from $t_{org}=0$. Multiple source nodes exist in the network $s_i$ while only one destination $d$,i.e. the WSN sink node is present. For each packet stream $p_j$ with average data rate $r_j$, there is associated a pair of destination node and source node $(s_{i,j},d)$.It describes the packet stream $p_j$ routed to the sink node $d$ from source node $s_i$. Let the number of available routes to choose be a constant number $K$ for all the packet streams, an approach similar to \cite{volvwagen}. For each route $route_{j,k}$, there exists a set of edges ${edge_m}$ that connect with each other to compose the route.$route_{j,k}$ describes the $k^{th}$ route for packet stream $j$.$K$ is the maximum number of routes available for each packet stream to be routed from its source node to the destination node.

At a given moment $t' \in \{n\Delta t,(n+1)\Delta t\}$, let there be \textit{L} packet streams. Consequently there would be a binary set ${x_i}$ of size $K*L$ for $x_i$ represents the $i \pmod{K}$th optional route for path n= $\lfloor \frac{i}{k} \rfloor$. $x_i \in \{0,1\}$.
The edges associated with $x_i$ is ${edge_{i,j}}$. For each edge \textit{j}, the length is $l_j$.

\subsection{Example Illustration}

\begin{figure}
    \centering
    \includegraphics[width=\linewidth]{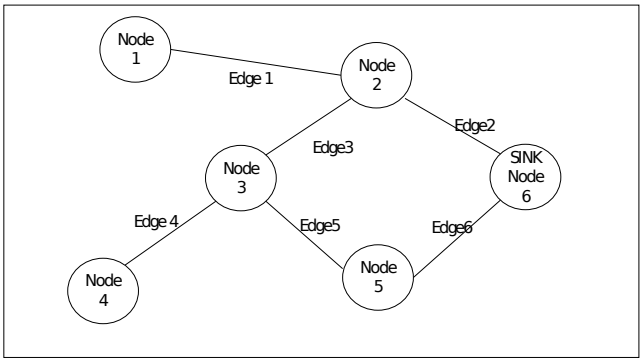}
    \caption{A small network example}
    \label{fig:1}
\end{figure}
\FloatBarrier

There are five source nodes and one sink in Figure \ref{fig:1}. Suppose only node $1$ and node $3$ are transmitting at the moment of $t$, each with an average data rate of $r_1$ and $r_3$.
For node $1$ to sink $6$, there are two paths $p_{1,1}$ and $p_{1,2}$. For node $3$ to sink $6$, there are also two paths $p_{3,1}$ and $p_{3,2}$.
From the graph we realise that $p_{1,1}=[edg_1,edg_2]$, $p_{1,2}=[edg_1,edg_3,edg_5,edg_6]$, $p_{3,1}=[edg_3,edg_2]$ and $p_{3,2}=[edg_5,edg_6]$.$edg_i$ is the $i^{th}$ edge.
Set $\bar{X_1}=[xp_{1,1},xp_{1,2}]$ and $\bar{X_2}=[xp_{3,1},xp_{3,2}]$, where $xp_{i,j}\in \{0,1\}$ indicates whether the \textit{j}th path for traffic stream starting from node \textit{i} is selected or not. 
For each path, we assume the energy consumption per interval $\triangle t$ is $e_{i,j}=\sum_{edg_k} e_{i,j,k}$, where $e_{i,j,k}$ is the energy consumption at $k^{th}$ edge with source number $i$ whose path index is $j$.

The overall energy consumption is $E=xp_{1,1}e_{1,1}+xp_{1,2}e_{1,2}+xp_{3,1}e_{3,1}+xp_{3,2}e_{3,2}$, where $xp_{1,1}+xp_{1,2}=1$ and $xp_{3,1}+xp_{3,2}=1$.

The conditions indicate for each traffic stream, only one path can be selected.

There exists another condition, which is the edge load should not exceed the maximum capacity. Assume for the edges within the network, the maximum capacity is uniform as $C_{max}$.
As a simple combinatorial problem, it is easy to deduce that there will be 4 combinations. Take one combination [$\bar{X_{1,1}}$,$\bar{X_{3,1}}$], for example, there are three edges in use.For $edg_1$ and $edg_3$, the edge load is $r_1\le C_{max}$ and $r_3 \le C_{max}$ respectively. For $edg_2$, the edge load is $r_1+r_3 \le C_{max}$.

\subsection{Energy Model}
 In this work, we use the energy model from \cite{energymodel}. 

\begin{equation}
E_{Tx}(l,d)=\begin{cases}
lE_{elec}+l\epsilon_{fs}d^2, & d < d_0.\\
lE_{elec}+l\epsilon_{mp}d^4, & \text{else}
\end{cases}
\end{equation}

The second item in the formula is the transmission energy for $l$ bits and the first item is the device holding energy for $l$ bits.$d$ is the distance of the edge connecting the transmitter and the receiver. 
\begin{equation}
E_{Rx}(l,d)=lE_{elec}
\end{equation}

\begin{table}[h]
\caption{Parameter Value}
\centering
\begin{tabular}{c r}
\hline\hline
Name&Value\\
\hline
$E_{elec}$& 50 nJ/bit\\
$\epsilon_{mp}$& $0.0013 \frac{pj}{bit}/m^4$\\
$\epsilon_{fs}$&$10 \frac{pj}{bit}/m^2$\\
$d_0$&$\sqrt{\frac{\epsilon_{fs}}{\epsilon_{mp}}}=87.7058$\\
\hline
\end{tabular}
\label{tab:hresult}
\end{table}

\subsection{Objective Function}
The objective function is the summation of energy consumption per path per edge per computation interval subject to bandwidth capacity and the encoding format according to domain wall encoding ~\cite{nick}.

Suppose $f(d_j)$ is the transmission power consumption on edge $j$ with length $d_j$.Suppose $\Bar{X}={x_i|i \le K}$ indicates whether a path has been selected or not. $K$ is the total number of available paths. We go over all the edges and within each edge $j$, we go through all the paths.And within each valid path $i$ that is indicated by $x_i$, we add the edge power consumption which includes both the transmitter and receiver power consumption.
\begin{equation}
\begin{aligned}
\min_{x_i} \quad & \sum_{all\ edge \ j} \{f(d_j) (\sum_{j \in x_i} x_ir_n\Delta t) + 2*E_{elec}*\sum_{j \in x_i}x_ir_n\Delta t\}\\
\textrm{s.t.} \quad & \sum_{j \in x_i} x_i*r_n \le C_{max} \   \forall edge_j\\
  & \sum_{x_i \in n} x_i=1 \ \forall n \\
\end{aligned}
\end{equation}
We apply slack variable technique to mitigate the inequality and equality constraint. As $x_i$ is either $0$ or $1$, it can be equivalently transferred to $x^2_i$ such that the objective function becomes a quadratic function with a constant term, which will be omitted in computation.
\begin{equation}
\begin{aligned}
f_{obj}=& \sum_{all\ edge\ j} \{f(d_j)(\sum_{j \in x_i}x_ir_n\Delta t)+\\
&2*E_{elec}\sum_{j \in x_i}x_ir_n\Delta t + \lambda_{1}(\sum_{j \in x_i} x_ir_n-C_{max})\} +\\
&\sum_{all\ stream\ n}\lambda_{2} (\sum_{x_i \in n} x_i-1)^2\\
=& \sum_{all\ edge\ j}\{\sum_{j \in x_i}(x_i)^2[r_n\lambda_{1}+2*E_{elec}r_n\Delta t + f(d_j)r_n\Delta t]\}-\\
&C_{max}*\lambda_{1}*\sum_{all\ edge\ j} \sum_{all\ stream \ n} \bold{I}_{j,n} \\
          & +\sum_{all\ stream\ n}{(\lambda_{2}-2){\sum_{x_i \in n}(x_i)^2}+2\lambda_{2}\sum_{i \neq j}x_ix_j} +\\
          &\lambda_{2}*n\\
\end{aligned}
\end{equation}

As the second constraint $sum_{x_i \in n} x_i=1 \forall n $ of the optimisation problem falls into the one hot encoding format. In order to save the computational resources (number of logical/physical qubits) so as to ease the actual computational process, we converted the encoding of the whole problem into domain wall encoding, the explanation of which is as follows.
\section{Translation to Domain-Wall Encoding}

Motivated by the enhanced performance seen in \cite{Chen21a,Berwald21a} (although not yet shown on a real engineering problem), we employ the domain-wall encoding scheme first proposed in \cite{nick}, translate to this encoding we first replace one-hot constraints with:
\begin{equation}
H_{K}=-\lambda[\sum_{i=0}^{K-2}Z_iZ_{i+1}-Z_0+Z_{K-1}]
\end{equation}

Where $Z_i=-2\bar{x}_i+1$ are Ising variables used to construct that encoding (note that the $\bar{x}$ is used to distinguish these variables from the original variables $x$), since we wish to work in a QUBO formulation , we substitute to obtain,
\begin{equation}
H_{K}=-\lambda[\sum_{i=0}^{K-2}4\bar{x}_i\bar{x}_{i+1}-2\bar{x}_i-2\bar{x}_{i+1}+2\bar{x}_0-2\bar{x}_{K-1}] -\lambda
\end{equation}
which can directly be used to replace variables under a one hot constraint such that
\begin{equation}
    x_i\rightarrow \begin{cases} \bar{x}_i & i=0 \\ \bar{x}_{i}-\bar{x}_{i-1} & 0<i<K-1 \\ -\bar{x}_{i-1} & i=K-1 \\ \mathrm{undefined} & \mathrm{otherwise} \end{cases},
\end{equation}
since this translation takes linear terms to linear terms, a quadratic formula in $x$ will also be be quadratic in $\bar{x}$






\section{Algorithm}

\begin{algorithm}
\DontPrintSemicolon
\SetAlgoLined
\KwResult{The Mapped Matrix Q}
\SetKwInOut{Input}{Input}\SetKwInOut{Output}{Output}
\Input{Graph}
\Output{Q}
\BlankLine
\ForEach{$edge_j$}{    
    
    \BlankLine  
    \ForEach{each original $route_i$}{
        \If{$route_i$ go through $edge_j$ }{
          $Q[i,i]=Q[i,i]+[`{r_n}^2*(\lambda_{1}-2\lambda_2C_{max})+2E_{elec}r_n\Delta t + f(d_j)r_n\Delta t]$\\
          \ForEach{$route_h$ go through $edge_j$ and $h \neq i$}
          {
          $Q[i,h]=Q[i,h]+\lambda_12{r_n}^2$
          }
        }
                              
    }    
    
    \ForEach{$ KL+(j-1)K'<i \le KL+jK'$}{
    	$k= (i-KL)\bmod K'$\\
        $Q[i][i]=Q[i][i]+(\lambda_1k^2-2\lambda_1C_{max}k)$\\
    	\ForEach{$i<q<KL+jK'$ }{
		$Q[i][q]=2\lambda_1k^2$\\
	}
    
    }

}
\BlankLine
\ForEach{each $packet stream_n$}{    
    
    \BlankLine  
    \ForEach{each $route_i$ belongs to $stream_n$}{
    $Q[i,i]=Q[i,i]+\lambda_{2}-2$ \\
    \ForEach{$route_j$ belongs to $stream_n$ and $j > i$}{
    $Q[i,j]=Q[i,j]+2*\lambda_{2}$
    
    	}
   
        }
                              
}
\caption{Mapping Algorithm}
\end{algorithm}
\begin{algorithm}
\caption{Hybrid Algorithm Procedure}
\begin{algorithmic}[1]
\STATE Call the sub procedure to collect all feasible paths - PathCollector
\STATE Call the sub procedure to assign paths to respective edges - getEdgeM \label{lst:line:2}
\STATE Call the sub procedure to formulate QUBO problem - makeEffArray
\STATE Call the sub procedure to encode the QUBO problem - makeEncoding
\STATE Call the QPU API Solver
\end{algorithmic}
\label{algo2}
\end{algorithm}

\begin{algorithm}
\caption{Path Collector}
\begin{algorithmic}[2]
\REQUIRE adjacency matrix, destination ID, source ID array,maxflownum
\FORALL{destination and source pair}
\STATE no node has been assigned the relay role yet
\WHILE{path amount is less than maxflownum}
\WHILE{the last relay node is not the destination node}
    \STATE Find the next relay node
    \STATE Tick this node as assigned
\ENDWHILE
\ENDWHILE
\ENDFOR
\end{algorithmic}
\end{algorithm}

Sub procedure $getEdgeM$ in line ~\ref{lst:line:2} of algorithm ~\ref{algo2} is to form a virtual three dimensional matrix that $pID=M_{i,j,k}$. $pID$ is the path ID assigned consecutively when running the $Path Collector$ algorithm. In $M_{i,j,k}$,$i,j$ is the node ID and $(i,j)$ indicates the edge that connects node $i$ and node $j$. $k$ indicates the $k^{th}$ path that goes through the edge $(i,j)$. In implementation, a two dimension matrix of size $N*N$ by $Path\_Amount$ is created instead for manipulation convenience. In this case, $(i,j,k)$ in the virtual three dimensional matrix corresponds to the ${((i-1)*N+j)}^{th}$ position in row and $k^{th}$position in column in the real two dimensional matrix. 
\section{Experiment}
The goal of the experiment is to evaluate the performance of the QPU against classical solvers (Cplex and Gurobi) in multi-objective routing problem that has been formulated into a QUBO. We expect that QPU given current hardware maturity can guarantee a solution quaility as good as the classical solvers whilst at a faster speed.
\begin{table*}[ht]
\caption{The list of parameter configurations used in this paper}
\centering
\begin{tabular}{p{0.55\linewidth}p{0.45\linewidth}}
\hline
Parameters & Configurations \\
\hline
$C_{max}$       &   5 \\
$\max r_{j}$.   &  5 \\
$r_{j}$   & follows uniform distribution\\
annealing time & 20 $\mu$s\\
number of samples per run & 10\\
$\max graph\ size$ & 12\\
$\min graph\ size$ & 4\\
number of problem instances per graph & 20\\

\hline
\end{tabular}
\end{table*}
\subsection{Configuration and Set-up}
Two random number generators are used. One is to generate the probability following uniform distribution. That is $p$. If $p>0.5$, a value is generated by the second random number generator uniformly distributed over 1 to 5 and it is assigned to be the corresponding flow rate.

Since we are interested in the ability of the annealers to provide samples very quickly, we have used $10$ samples for every anneal, except where stated otherwise. For the classical solvers, we deployed timers to record the time before the solver call and after the solver call and calculate the lap between them. For the QPU, we use qpu\_sampling\_time within the 'timing' info as the processing time per run. We are using the default anneal time of 20 $\mu$s for each run. We further use the fixed\_variable technique \cite{fixedvariable} to slim the effective QUBO size submitted to the QPU. 

There are two target experiment types that we have done. The first is that we go over all  the possible combinations of graph size up to size 12 and source number excluding the randomness of the flow rate per source and the second is that we apply Erdos-Reny{\'i} graph generation algorithm (assigning different edge existence probabilities) and generate 20 problem samples each graph size from (4 to 12) and run the statistical analysis.

\subsection{Experimental methods and data reporting \label{sec:meth_and_report}}
There are several quantities which we used as axes on our plots, for the readers convenience we define them here (see section \ref{sec:prob_form} for details of problem formulation):

\begin{itemize}

\item Source number: the number of nodes on the network graph which act as sources within the network.

\item Graph size: the size of the graph used in the network problem, larger graphs will typically lead to larger QUBOs.

\item Edge probability: the probability of an edge appearing within an Erd{\"o}s Renyi random graph used to construct the routing problems.

\item QUBO size: the number of binary variables used in the problem which is passed to the annealer or classical solver, this number is always quoted before minor embedding is performed to allow a fair comparison across solvers. The QUBO size we report is after variable fixing has been performed.

\item Performance degradation graph size: the point along axis one step ahead of the cross point where Correct Rate intersects with Incorrect Rate or Embedding Error Rate.

\item Correctness rate: the number of solutions from QPU that reach the minimum to the overall number of problem instances (the fraction of the samples which returned an optimal solution).

\item Processing time: the time taken to attain a feasible solution

\item Embedding error rate: the number of problem instances that fail to be embedded onto the QPU to the overall number of problem instances.

\item Incorrect rate: the number of solutions from the QPU where none of the returned solutions reach the minimum energy found for all solvers. Note that Correctness rate +Incorrect rate+ Embedding error rate sum to one.

\end{itemize}

We often plot the correctness rate for different solvers and we use this as a key metric to compare performance between the QPUs and classical solver over different quantities such as QUBO size, source number or the graph size.

We used macOS Sierra version 10.12.6 to run the classical algorithms. The processor is 3.4GHz Intel Core i7 and the memory is 16GB 1333MHz DDR3.

For all experiments reported here we performed $10$ reads on the quantum annealer, we have chosen this relatively small number of reads to assess the ability to attain a ``good solution quickly'' as it is likely that solving network problems like those described here will be very time-sensitive in most real applications. It is likely that some quantities, such as the probably that a valid solution is ever found for a given problem, would be substantially improved by taking more reads.

\section{Results}

\begin{figure}[htp]
    \centering
    \includegraphics[width=\linewidth]{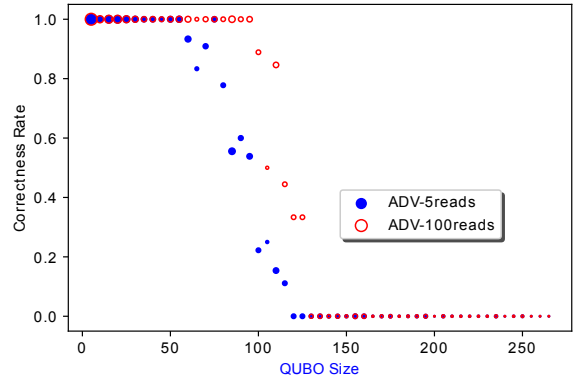}
    \caption{Advantage\_System Correctness Plot \_ Erdos\_Renyi: Plot of correctness rates for graph sizes up to size \textcolor{red}{12} and source numbers one up to \textcolor{red}{11}, and edge probablities \textcolor{red}{0.6}- \textcolor{red}{0.9}.  
    }
    \label{fig:size_correctness}
\end{figure}

Before getting into discussion of how the plots depend on the properties of the graph problem we are solving, it is worth briefly stepping back and seeing how they depend on lower level properties, from figure \ref{fig:size_correctness}, we see that for a broad sampling of the data, the correctness rate correlates strongly with QUBO size with a relatively narrow window where success probability drops off from approximately $1$ to approximately $0$. We also observe that the the success probability is increased by taking more reads, however in this paper we are interested in being able to get a good solution very quickly, so we base the data in the remainder of the paper on $10$ reads.
\subsection{Correctness}
\subsubsection{Correctness Rate based on brute force graph generation algorithm}
\begin{figure}[htp]
    \centering
    \includegraphics[width=\linewidth]{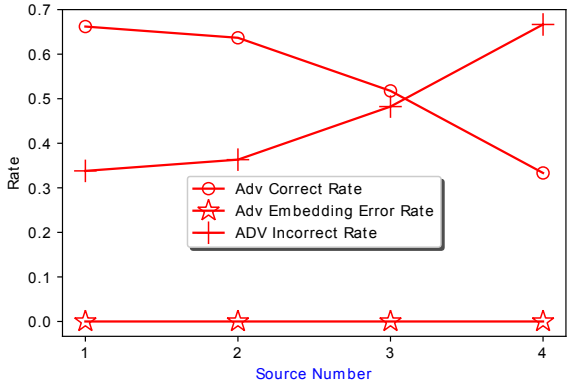}
    \caption{Plot of different measures of success for Advantage\_System QPU versus source number using exhaustively generated graphs of size $5$, see text for details. }
    \label{fig:adv_size_5_exhaust}
\end{figure}

\begin{figure}[htp]
    \centering
    \includegraphics[width=\linewidth]{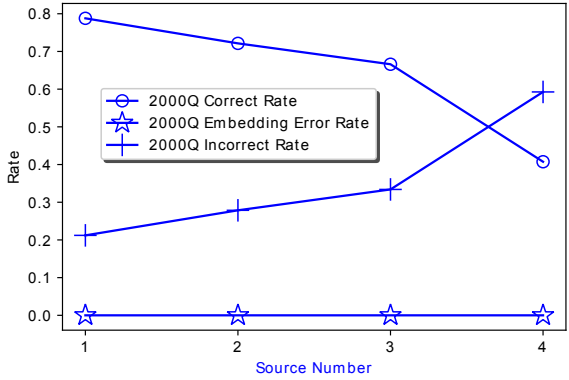}
    \caption{Plot of different measures of success for 2000Q QPU versus source number using exhaustively generated graphs of size $5$, see text for details. }
    \label{fig:2000Q_size_5_exhaust}
\end{figure}

We ran over all the possible combinations of graph topology and traffic source for graph size $4$ and $5$.
The plots in figs. \ref{fig:adv_size_5_exhaust},\ref{fig:2000Q_size_5_exhaust} show the correctness rate (defined as the fraction of feasible solutions obtained in a given experiment) against the source number. There are three categories that the result data from QPU can fall into: 1. QPU reports embedding error and hence can not solve the problem; 2. QPU solves the problem but it is not the most optimal; 3. QPU provides the most optimal solution.  We note that for all cases at graph size 5, the the problem can be embedded and solved to optimality by the QPU.

 We also note from fig.~\ref{fig:adv_Gurobi_time_scatter} that if more samples were taken the range in which this transition occurs increases, but the qualitative shape remains the same. In this paper we are restricting ourselves to studying how the annealer performs when restricted to running for a very short time and therefore a small number of samples, but it is worth remarking that many of the problems tested in this paper could be solved eventually with more anneals.



 From these plots, we can tell from the data that advantage\_sys1.1 has an absolute advantage in terms of speed as QUBO size increases to around $10$. Up to around size $20$, the QPUs are typically able to solve the problems within the $10$ reads we take. By a one-by-one eye-check of the data, we can tell that in these experiments 2000Q also outperforms classical solvers in terms of speed. As the solution is not found as quickly as the advantage\_sys1.1, in this plot, the speediness correctness rate (the ratio of the fraction of cases where a solution is found faster) for 2000Q is always 0.


Figures \ref{fig:adv_size_5_exhaust},\ref{fig:2000Q_size_5_exhaust} depict results for problems generated exhaustively for graph size $5$ with source numbers $1-4$ and all possible connected graphs. The measures are explained in section \ref{sec:meth_and_report}, the correct and incorrect rate refer to cases where the most optimal solution was or was not found respectively. The embedding error rate refers to cases where the problem could not be embedded successfully.
We can tell from these figures that the correctness rate decreases as the source number increases for both advantage system and 2000Q.  This is probably because the QUBO size increases when the number of sources increases. Furthermore, advantage system keeps a more than $60\%$ faster rate than both the classical solvers and 2000Q across all possible number of source nodes. Even further, at the highest number of source nodes, the faster rate for advantage system reaches the highest value among all. We suspect it is because given highest number of source nodes, the practical QUBOs submitted for all the four solvers become more complicated. While classical solvers can tackle smaller size of QUBO problem at a faster rate (we suspect if the QPU processing time might decrease at the same solution quality if we reduce the sample number per run to $5$ or $3$), they are losing ground of larger QUBO size to QPU solvers possibly because of the parallel solution searching mechanism facilitated by quantum mechanics. 

From the analysis for data collected from all the problem instances for graph size $4$ that are feasible for submission to the QPU, we can tell that the QPU demonstrates an absolute advantage over the classical solvers we tested in terms of  solution quality across all the problem space. It doesn't show an advantage in the speed with number of reads equal to 3000. It is because the overall QUBO size is small ($= 5$) so it is fast enough for classical solvers to attain a correct solution and when the number of runs is to be decreased to $5$, the QPU speed will be overriding those by the classical solvers without degrading the solution quality.



\FloatBarrier
\subsubsection{Correctness Rate based on probabilistic graph generation algorithm}
\begin{figure}[htp]
    \centering
    \includegraphics[width=\linewidth]{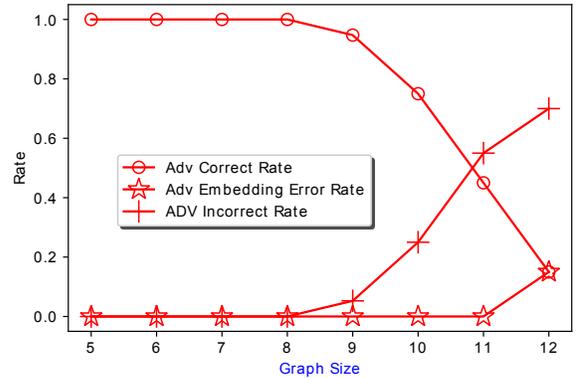}
    \caption{Plot of different measures of success for Advantage\_System QPU versus graph size for graphs generated with the Erdos-Renyi algorithm with an edge probability of $0.6$. }
    \label{fig:adv_p0.6}
\end{figure}

\begin{figure}[htp]
    \centering
    \includegraphics[width=\linewidth]{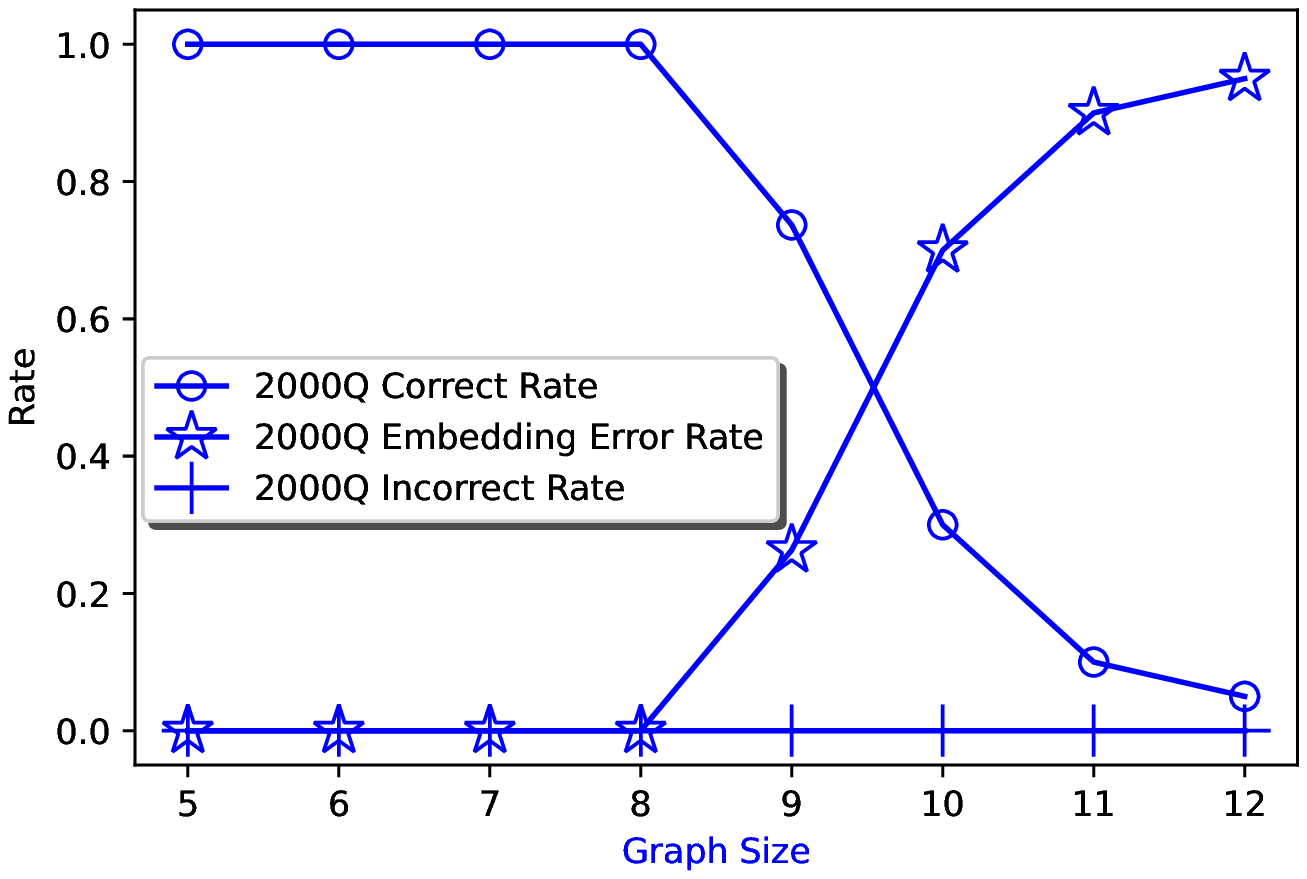}
    \caption{Plot of different measures of success for 2000Q QPU versus graph size for graphs generated with the Erdos-Renyi algorithm with an edge probability of $0.6$, see text for details. }
    \label{fig:2000Q_p0.6}
\end{figure}

\begin{figure}[htp]
    \centering
    \includegraphics[width=\linewidth]{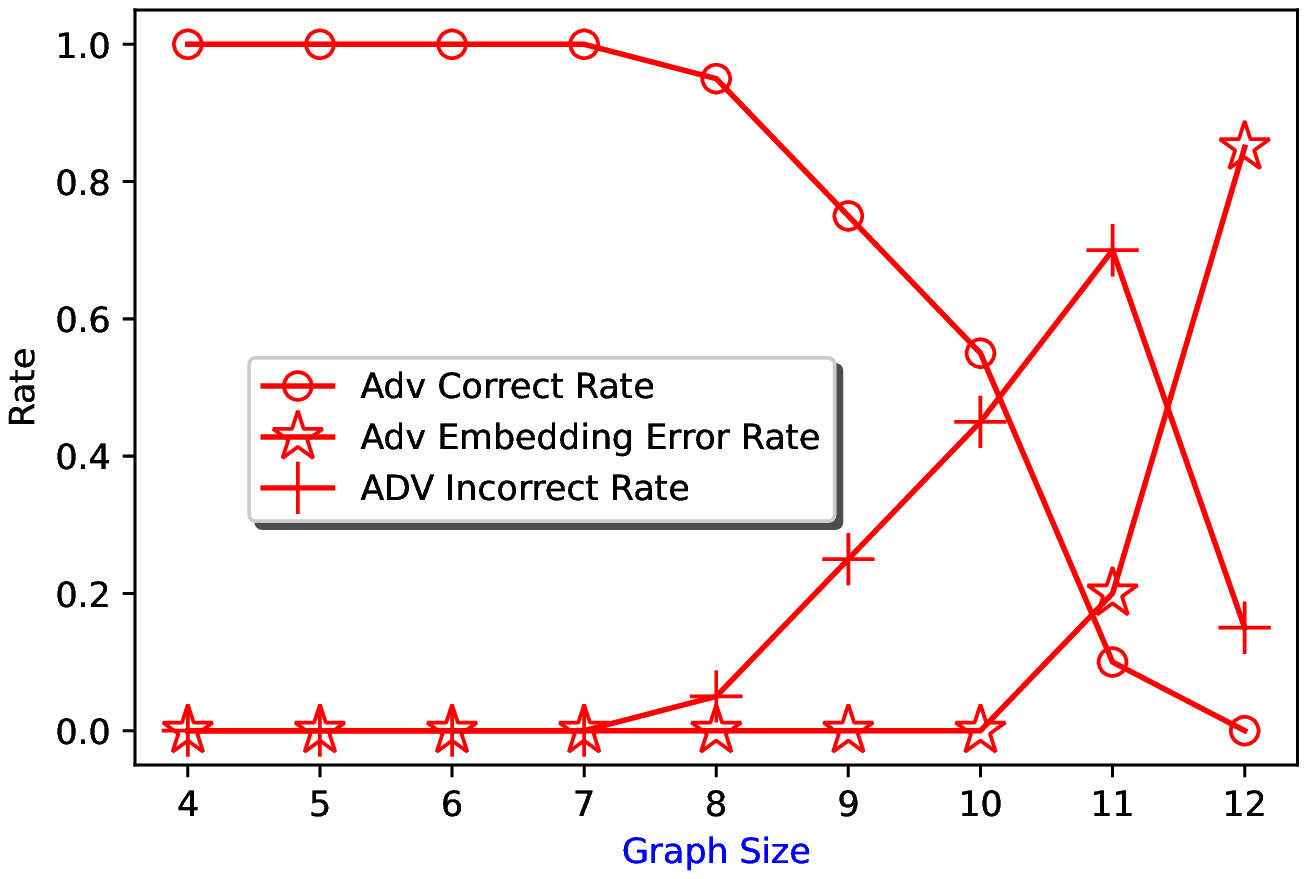}
    \caption{Plot of different measures of success for Advantage\_System QPU versus graph size for graphs generated with the Erdos-Renyi algorithm with an edge probability of $0.7$, see text for details.}
    \label{fig:adv_p0.7}
\end{figure}

\begin{figure}[htp]
    \centering
    \includegraphics[width=\linewidth]{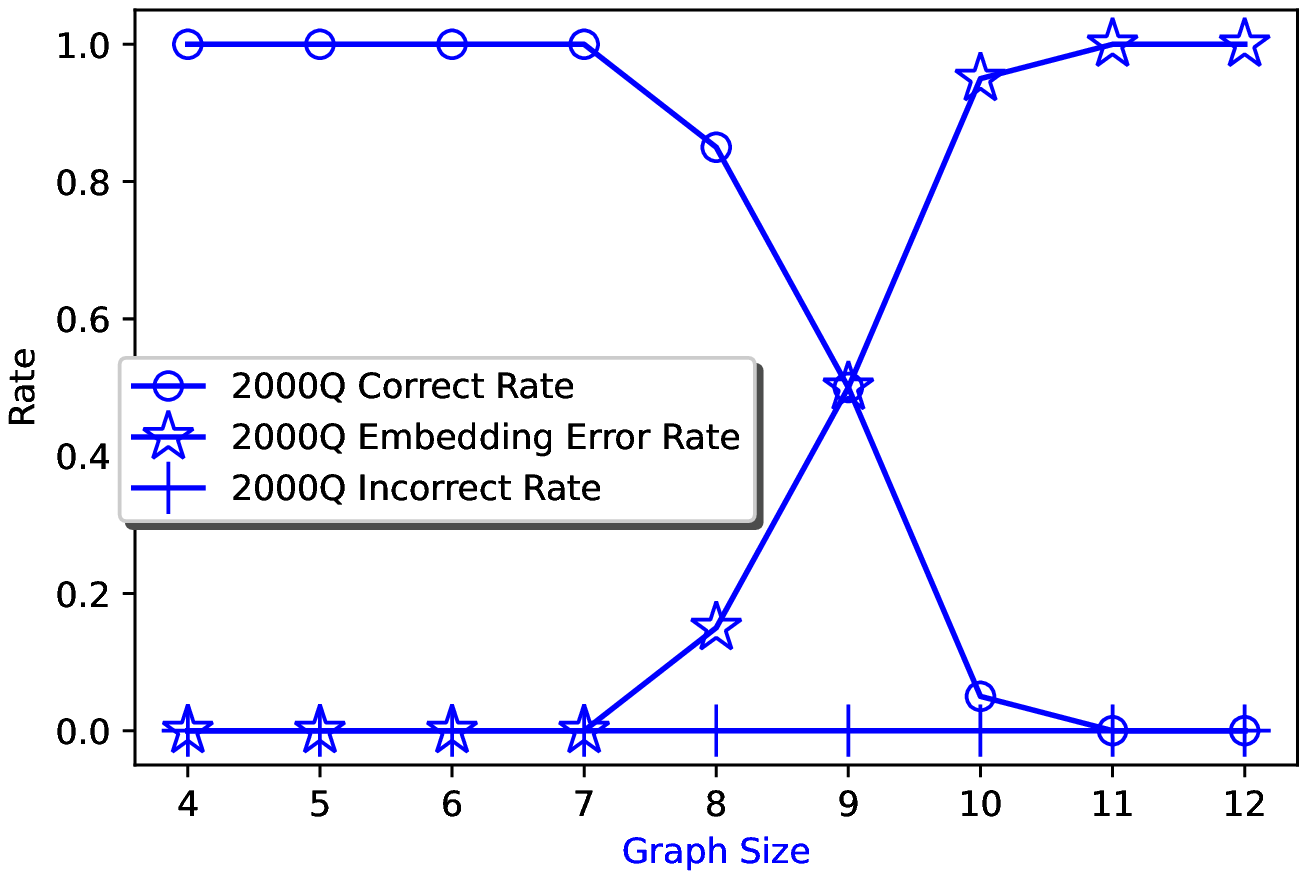}
    \caption{Plot of different measures of success for 2000Q QPU versus graph size for graphs generated with the Erdos-Renyi algorithm with an edge probability of $0.7$, see text for details.}
    \label{fig:2000Q_p0.7}
\end{figure}

\begin{figure}[htp]
    \centering
    \includegraphics[width=\linewidth]{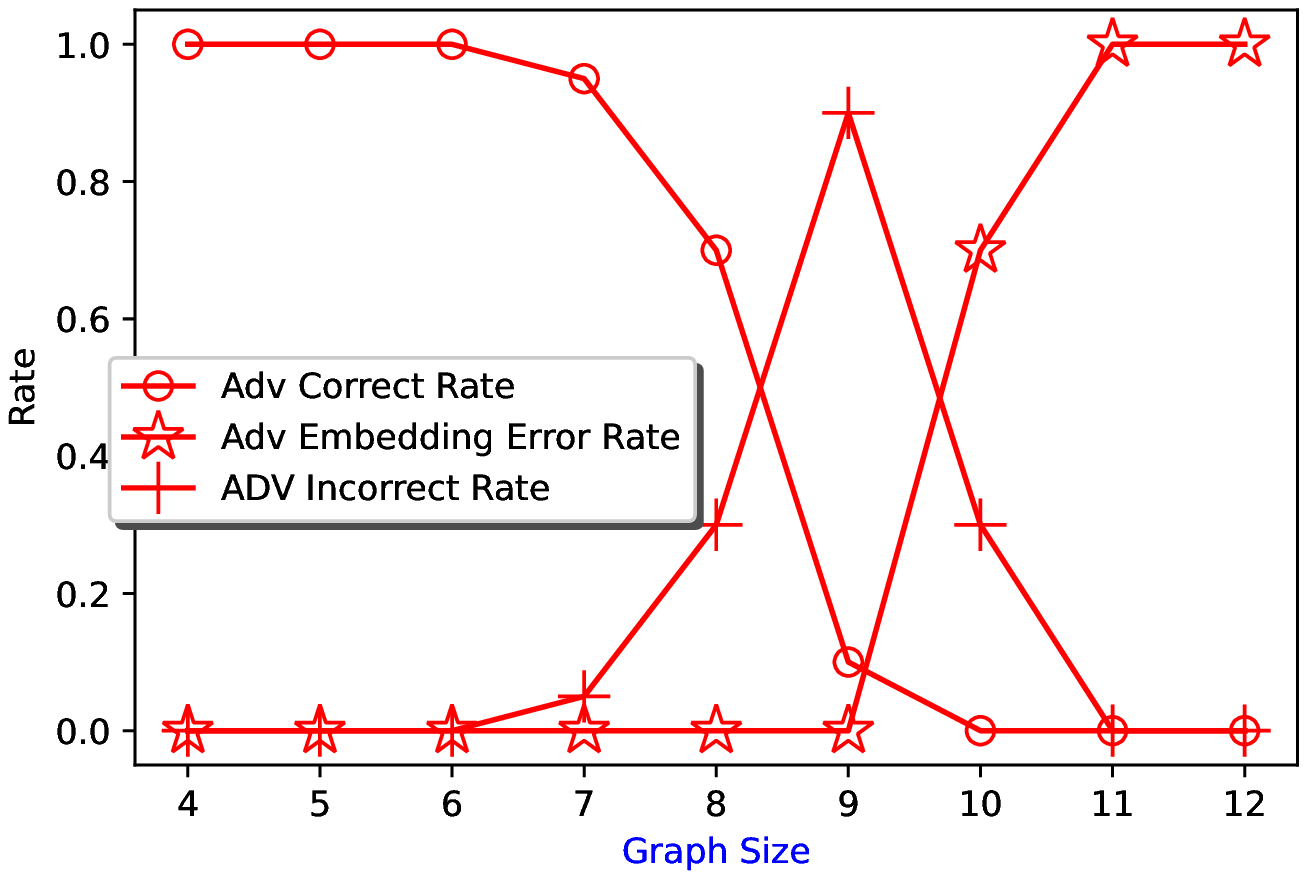}
    \caption{Plot of different measures of success for Advantage\_System QPU versus graph size for graphs generated with the Erdos-Renyi algorithm with an edge probability of $0.9$, see text for details.}
    \label{fig:adv_p0.9}
\end{figure}

\begin{figure}[htp]
    \centering
    \includegraphics[width=\linewidth]{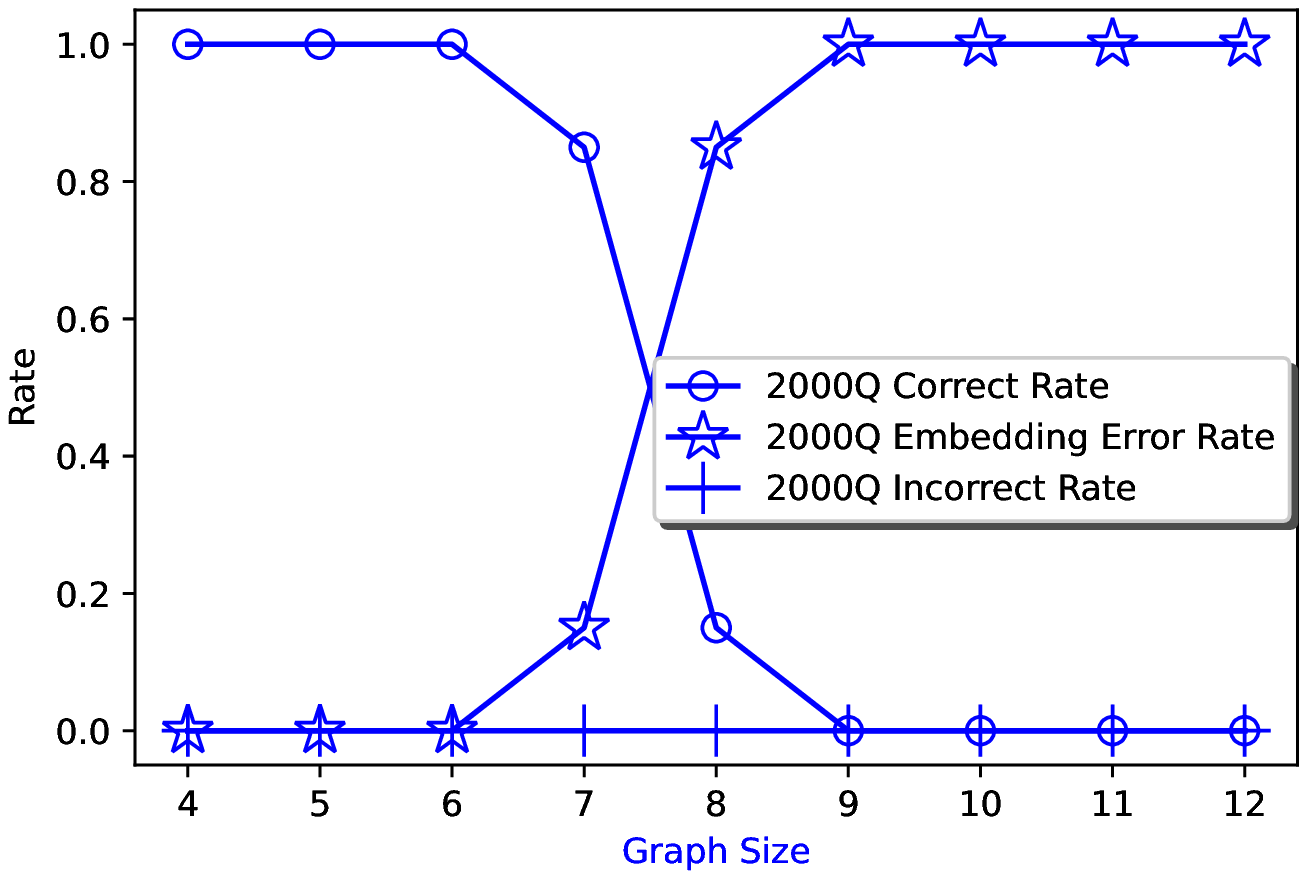}
    \caption{Plot of different measures of success for 2000Q QPU versus graph size for graphs generated with the Erdos-Renyi algorithm with an edge probability of $0.9$, see text for details.}
    \label{fig:2000Q_p0.9}
\end{figure}

\begin{figure}[htp]
    \centering
    \includegraphics[width=\linewidth]{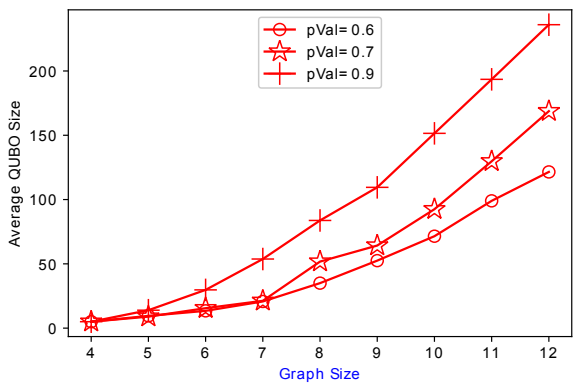}
    \caption{QUBO size versus graph size for graphs generated by the Erdos-Renyi algorithm for different edge probabilities.}
    \label{fig:size_edge}
\end{figure}

Figures \ref{fig:adv_p0.6},\ref{fig:2000Q_p0.6},\ref{fig:adv_p0.7},\ref{fig:2000Q_p0.7},\ref{fig:2000Q_p0.9},\ref{fig:adv_p0.9},\ref{fig:size_edge} depict relevant quantities for data collected by using the Erdos-Renyi graph generator with the edge probability set to $0.6$,$0.7$ and $0.9$ separately. We generated $20$ graph samples per edge probability and do the average in the analysis.
For figures \ref{fig:adv_p0.6},\ref{fig:2000Q_p0.6},\ref{fig:adv_p0.7},\ref{fig:2000Q_p0.7},\ref{fig:2000Q_p0.9},\ref{fig:adv_p0.9}, the correct and incorrect rate refer to cases where the most optimal solution was or was not found respectively. The embedding error rate refers to cases where the problem could not be embedded successfully, and the measures are explained in section \ref{sec:meth_and_report}. These plots differ only in the edge probability and the QPU used.

We can tell that for edge probability $0.6$, advantage\_system's performance starts to degrade at graph size $10$ while 2000Q system's performance starts to degrade at graph size $9$. Embedding errors are reported occasionally for graph size equal to $12$. 

As we increase the edge probability to $0.7$ ,intuitively we suspect the graph becomes more complex and \ref{fig:size_edge} shows that as the edge probability increases, the trimmed QUBO size increases. A larger share of embedding error cases emerges.For advantage\_system, the performance starts to degrade at graph size 9 while for 2000Q system, it is at graph size 8. Furthermore, more we notice that for advantage\_system, embedding errors occur by around $20\%$ for graph size equal to $11$ and this figure increases to $80\%$ when graph size reaches $12$. For 2000Q, embedding errors appear at graph size 8 at around $20\%$and reaches $100\%$ at graph size $11$ and onward. 

At the edge probability to $0.9$, the degradation graph size is 8 for advantage\_system while 7 for 2000Q. Furthermore, advantage\_system starts to report embedding error at higher rate when graph size reaches 10 and on wards while for 2000Q, the graph size value is 8 with around $80\%$ embedding error rate. 

\subsection{Speediness}
\subsubsection{Speediness based on brute force graph generation algorithm}
\begin{figure}[htp]
    \centering
    \includegraphics[width=\linewidth]{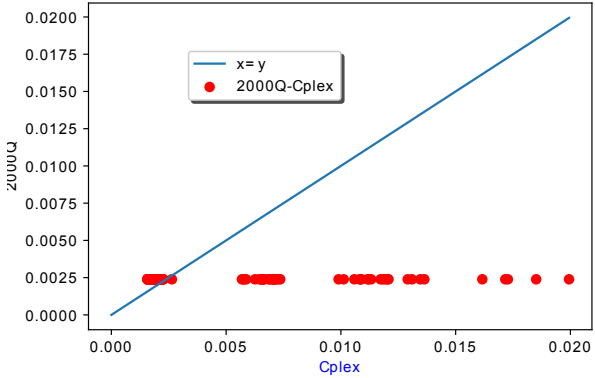}
    \caption{Scatter plot of speediness of 2000Q QPU versus Cplex. The blue line shows equal times an is a guide to the eye. Plotted for over all graphs of size 5. }
    \label{fig:scatter_2000Q_Cplex_5_exhaust}
\end{figure}

\begin{figure}[htp]
    \centering
    \includegraphics[width=\linewidth]{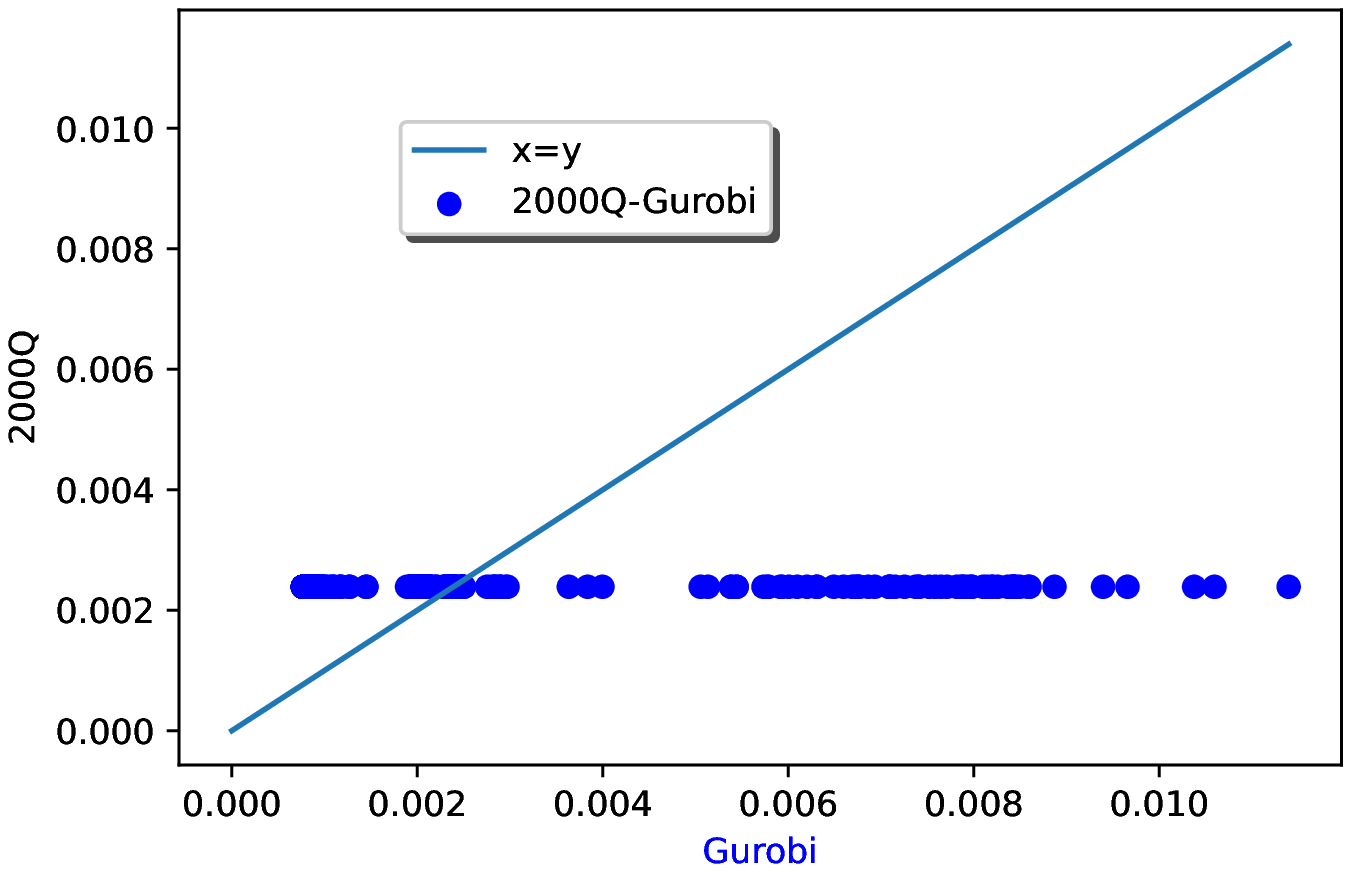}
    \caption{Scatter plot of speediness of 2000Q QPU versus Gurobi. The blue line shows equal times an is a guide to the eye. Plotted for over all graphs of size 5. }
    \label{fig:scatter_2000Q_Gurobi_5_exhaust}
\end{figure}

\begin{figure}[htp]
    \centering
    \includegraphics[width=\linewidth]{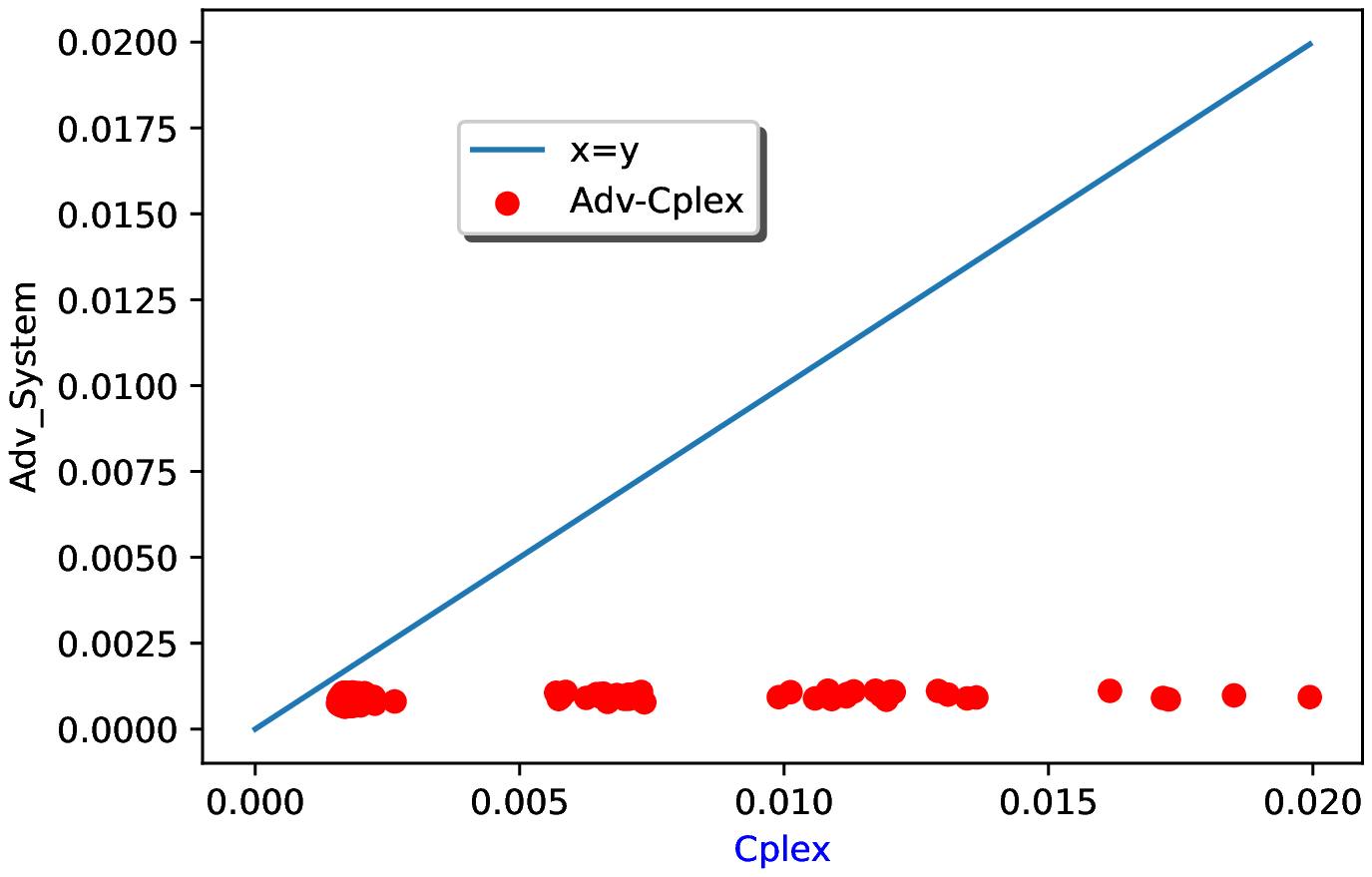}
    \caption{Scatter plot of speediness of Advantage\_System QPU versus Cplex. The blue line shows equal times an is a guide to the eye. Plotted for over all graphs of size 5. }
    \label{fig:scatter_adv_Cplex_5_exhaust}
\end{figure}

\begin{figure}[htp]
    \centering
    \includegraphics[width=\linewidth]{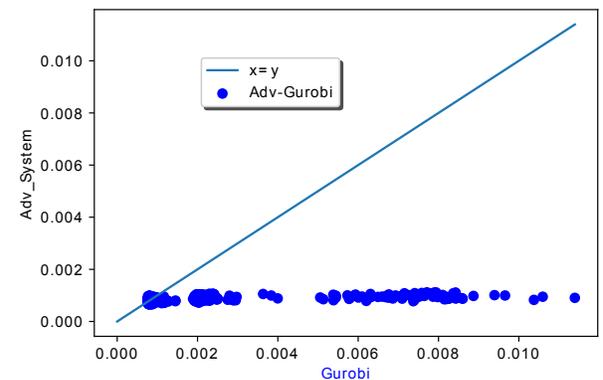}
    \caption{Scatter plot of speediness of Advantage\_System QPU versus Gurobi. The blue line shows equal times an is a guide to the eye. Plotted for over all graphs of size 5. }
    \label{fig:scatter_adv_Gurobi_5_exhaust}
\end{figure}

For network size of $5$, we compare the processing time between the classical solvers and the QPUs, we ignore cases where embedding errors were encountered for this analysis. The percentage of problems that reports embedding error is $57\%$. We observe from figures \ref{fig:scatter_2000Q_Cplex_5_exhaust},\ref{fig:scatter_2000Q_Gurobi_5_exhaust},\ref{fig:scatter_adv_Cplex_5_exhaust},\ref{fig:scatter_adv_Gurobi_5_exhaust} that QPUs have a constant value of processing time across almost all the valid problem instances. For 2000Q the processing time is $0.0025$ seconds while for Advantage\_system the processing time is nearly half that value. There exists a larger portion of valid problem instances for 2000Q that consume longer processing time that for Advantage\_system against both Gurobi and Cplex. We further notice that Cplex works noticeably longer than Gurobi for all valid problems.

\begin{figure}[htp]
    \centering
    \includegraphics[width=\linewidth]{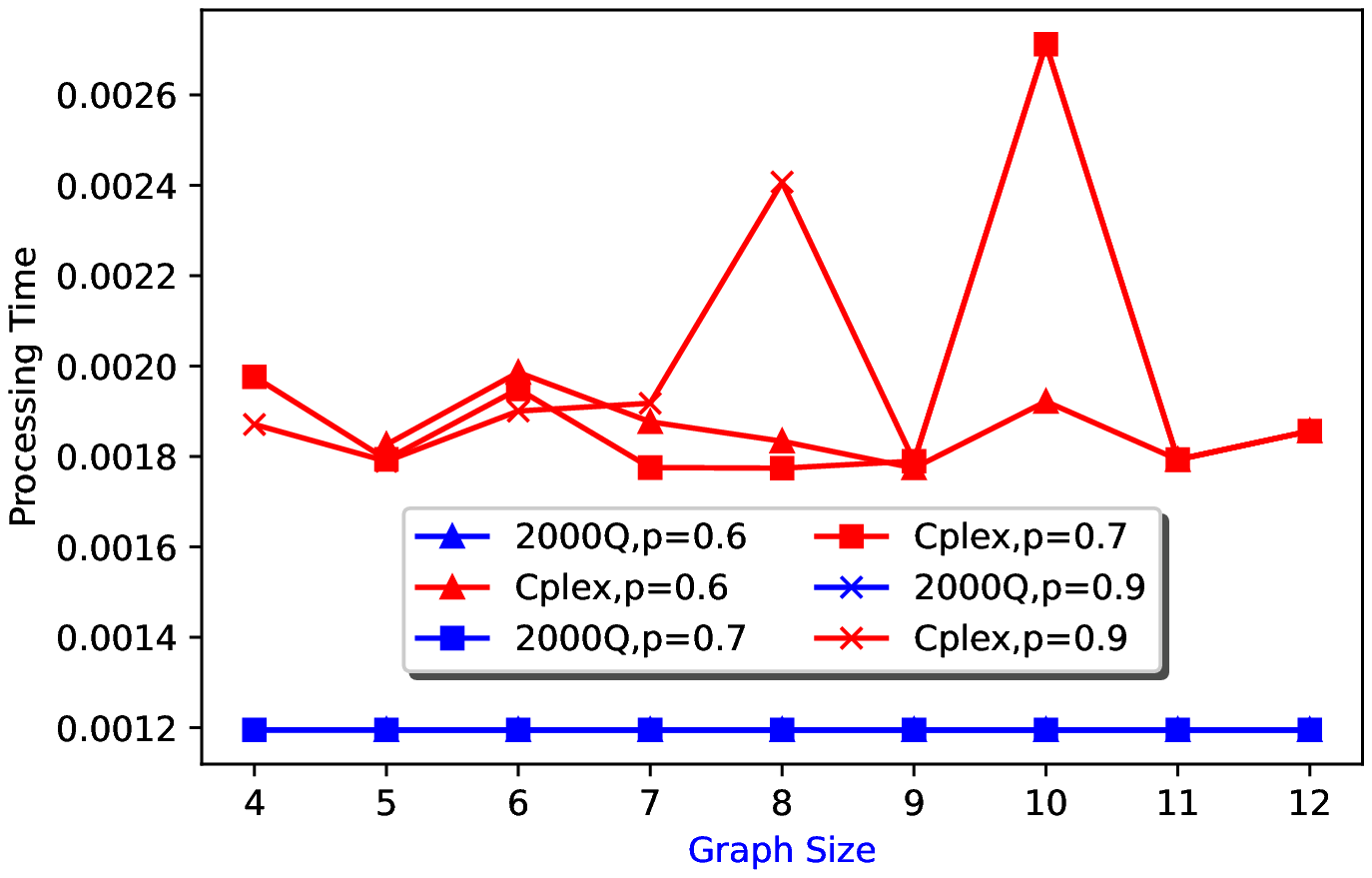}
    \caption{Comparison of 2000Q QPU Speediness with Cplex averaged for different graph sizes.This plot uses graphs generated with the Erdos-Renyi algorithm. }
    \label{fig:2000Q_Cplex_time_scatter}
\end{figure}

\begin{figure}[htp]
    \centering
    \includegraphics[width=\linewidth]{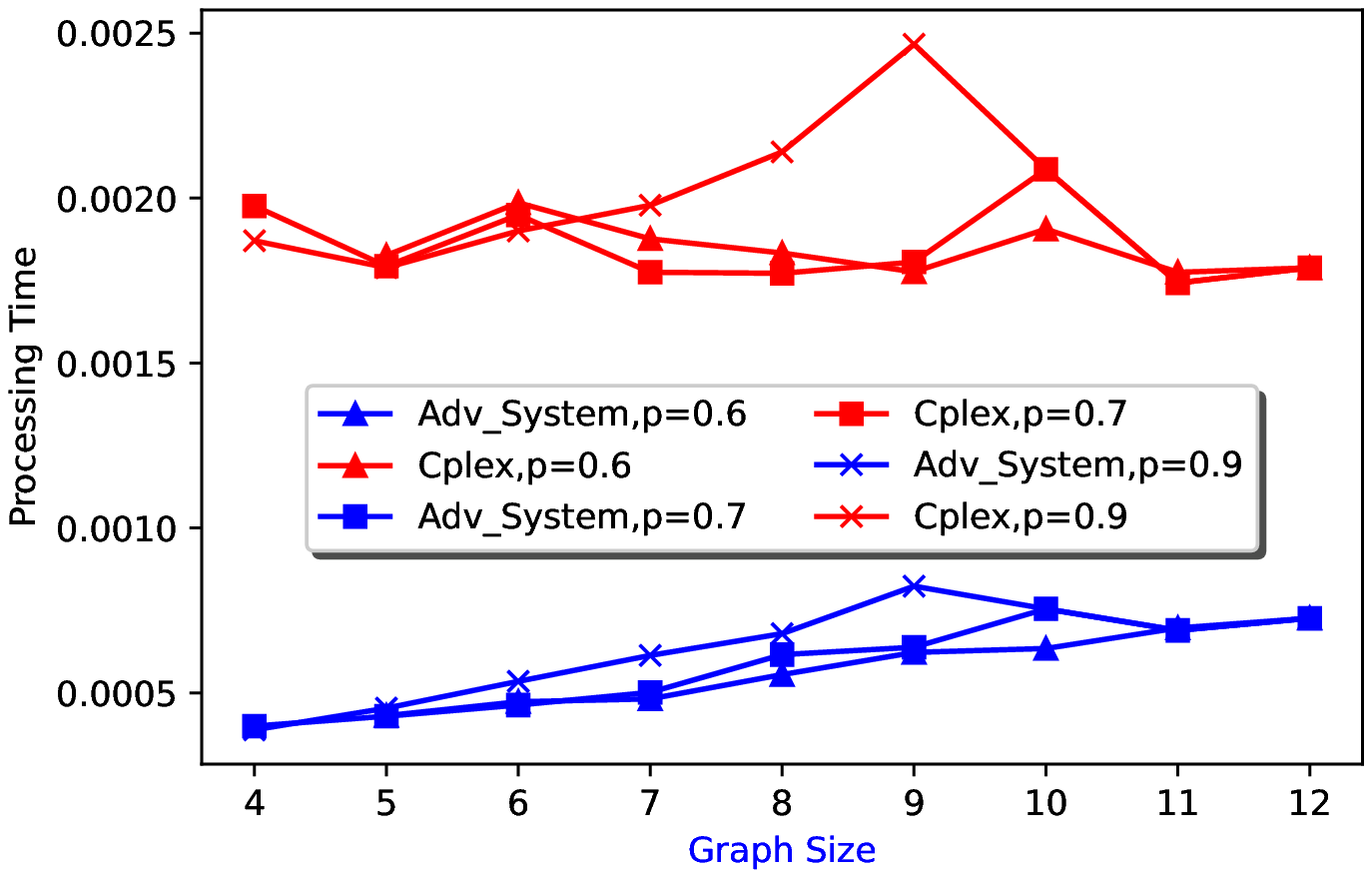}
    \caption{Comparison of Advantage\_System QPU Speediness with Cplex averaged for different graph sizes.This plot uses graphs generated with the Erdos-Renyi algorithm.  }
    \label{fig:adv_Cplex_time_scatter}
\end{figure}

\FloatBarrier
Figures \ref{fig:2000Q_Gurobi_time_scatter},\ref{fig:2000Q_Cplex_time_scatter},\ref{fig:adv_Cplex_time_scatter},\ref{fig:adv_Gurobi_time_scatter} show the average processing time (in seconds) per graph size for QPUs and classical solvers respectively. We excluded problems where embedding errors occurred on either QPU. Hence readers will notice that for figures \ref{fig:2000Q_Gurobi_time_scatter},\ref{fig:adv_Gurobi_time_scatter} and \ref{fig:2000Q_Cplex_time_scatter},\ref{fig:adv_Cplex_time_scatter}, the plots of the processing time for classical solvers have similar patterns respectively. To be more specific, from previous figures, we can tell that 2000Q is running faster to encounter embedding errors as the graph size increases, compared with the Advantage\_system. 
\subsubsection{Speediness based on probabilistic graph generation algorithm}
As the same with what has been observed from \ref{fig:scatter_2000Q_Cplex_5_exhaust}, 2000Q shows a constant processing time on average. Whilst for Advantage\_system, the average processing time goes up the graph size increases but stops increasing certain graph size and beyond. For Edge probability equal to $0.9$, this occurs at size $9$, for edge probability equal to $0.7$, this happens at size $11$ while for edge probability equal to $0.6$, the we cannot see this effect and hypothosize this may happen beyond size $12$ - our experiment graph size upper limit. From previous figures, we have come to agree that for edge probability equal to $0.9$ with graph size larger than $9$, only problem instances below graph size $9$ can consistently be embedded well onto the QPUs and return valid solutions. The similar conjecture can be applied to cases where the edge probability is less with larger turning points. For the phenomenon that average processing time is lessened for even larger graph size, we suspect the graph size alone can not determine the complexity of the network. Other factors take effect such as the structure of the network.

\begin{figure}[htp]
    \centering
    \includegraphics[width=\linewidth]{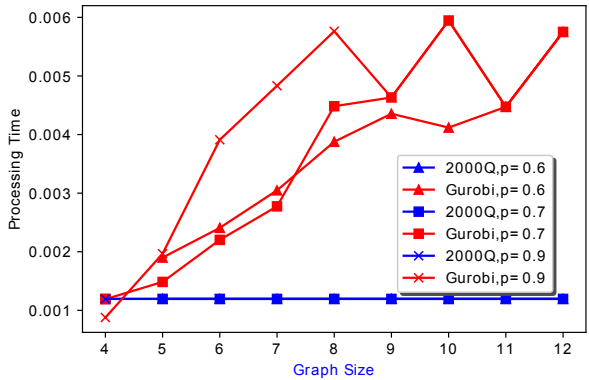}
    \caption{Comparison of 2000Q QPU Speediness with Gurobi averaged for different graph sizes.This plot uses graphs generated with the Erdos-Renyi algorithm. }
    \label{fig:2000Q_Gurobi_time_scatter}
\end{figure}

\begin{figure}[htp]
    \centering
    \includegraphics[width=\linewidth]{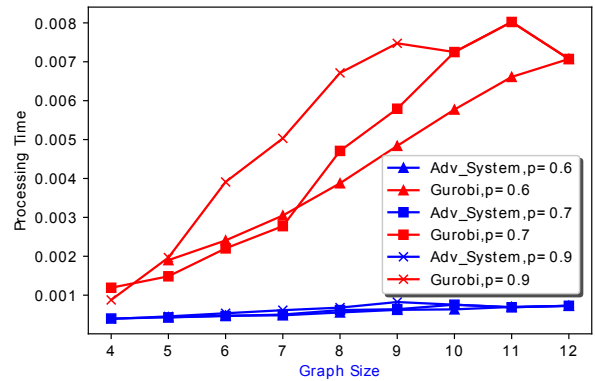}
    \caption{Comparison of Advantage\_System QPU Speediness with Gurobi averaged for different graph sizes.This plot uses graphs generated with the Erdos-Renyi algorithm.  }
    \label{fig:adv_Gurobi_time_scatter}
\end{figure}

The Cplex average processing time seems to correlate less with the graph size and the edge probability whilst for Gurobi, \ref{fig:2000Q_Gurobi_time_scatter},\ref{fig:adv_Gurobi_time_scatter} show that the average processing time goes up as the graph size increases in general and with larger edge probability. If this trend continues, it suggests that for larger problems there may be a crossover where Cplex becomes more effective.


\FloatBarrier

\section{Conclusion}

We have studied the performance of two D-Wave QPUs on small network routing problems with a limited number of reads from the processor. By comparing this to the well used Gurobi and Cplex classical solvers on a standard workstation, we find that both the 2000Q and Advantage yield superior performance in terms of absolute runtime. This is an encouraging proof-of-concept result for the application of similar QPUs to this kind of problem. We find that the most relevant quantity to determine QPU performance is the overall size of the QUBO which we apply it to, although we did find that both the size of the underlying network graph, and the number of sources had a significant effect as well. While most of the problems here involved small QUBOs, larger ones were occasionally also generated. Even within a few reads the QPUs were able to solve most QUBOs below size about $20$ and were not able to beyond this size. While this range is still accessible by exhaustive search methods, this study still provides useful proof-of-concept, as the problems could often be solved very quickly. This work suggests a route to practical quantum advantage where problems where problems which can be solved classically still yield an advantage by being solved much more quickly.

\subsection{Discussion}
Some preliminary experiments (not presented in this paper) are conducted in ns3, the codes of which has been made public via  https://github.com/cjie3331/quantumrouting. It demonstrated the feasibility of applying a QPU in the network routing design softwaredly. By experiments (presented in this paper) conducted in python, we have shown the advantage of QPU over classical solvers towards the same QUBO problem tailored to optimal route selection. It is our interest to run the performance comparison in ns3 between QPU and classical solvers in the near future and then compare the overall design to current state-of-art heuristic algorithm such as various localised learning algorithms. Due to the limited number of qubits supported by current QPU hardware, it is our plan to employ the clustering concept in sensor network to assign controller to each micro-network in a hierarchical manner. Last but not least, we intend to implement the overall hierarchical design into automated vehicular communication based on the justification that faster computation turn-around time can more seamlessly monitor/manage the communication process to its best effort.

\section{Acknowledgements}

This research used resources of the Oak Ridge Leadership Computing Facility, which is a DOE Office of Science User Facility supported under Contract DE-AC05-00OR22725.
Nicholas Chancellor and Jie Chen were funded by EPSRC fellowship EP/S00114X/1. The authors thank Adam Callison for useful discussions and contribution to the Python codes.The authors also acknowledge the support of CONNECT – the Science Foundation Ireland research centre for future networks and communications.

\bibliographystyle{unsrt}
\bibliography{bibliography}

\noindent

\clearpage

\end{document}